\documentclass[12pt]{article}
\usepackage{amsmath,amssymb,epsfig,color, url}
\usepackage{psfrag}
\usepackage{graphicx}
\usepackage{bm}

\newcommand{\be}{\begin{equation}}  
\newcommand{\ee}{\end{equation}}

\def\pslash{\rlap{\hspace{0.02cm}/}{p}}
\def\vslash{\rlap{\hspace{0.02cm}/}{v}}
\def\Dslash{\rlap{\hspace{0.07cm}/}{D}}

\begin{document}

\begin{titlepage}

\begin{flushright}
WSU-HEP-1502\\
March 24, 2015
\end{flushright}

\vspace{0.7cm}
\begin{center}
\Large\bf 
An Introduction to NRQED

\end{center}

\vspace{0.8cm}
\begin{center}
{\sc Gil Paz}\\
\vspace{0.4cm}
{\it 
Department of Physics and Astronomy \\
Wayne State University, Detroit, Michigan 48201, USA 
}

\end{center}
\vspace{1.0cm}
\begin{abstract}
  \vspace{0.2cm}
  \noindent
We present a pedagogical introduction to NRQED (non-relativistic quantum electrodynamics). NRQED is an effective field theory that describes the interaction of non-relativistic, possibly composite, spin-half particle with the electromagnetic field.  We explain in detail how the NRQED Lagrangian is constructed up to and including order $1/M^2$, where $M$ is the mass of the spin-half particle. As a sample application, we derive the Thomson scattering cross section for the low energy scattering of a photon and a possibly composite spin-half particle.   
\end{abstract}
\vfil

\end{titlepage}

\section{Introduction}
Non-relativistic quantum electrodynamics (NRQED) is an effective field theory that describes the interaction of a non-relativistic spin-half particle, possibly composite, with the electromagnetic field. 
NRQED was first introduced in a paper by Caswell and Lepage \cite{Caswell:1985ui} but its origins can be traced to the early days of quantum mechanics. Indeed, a standard course in non-relativistic quantum mechanics \cite{Cohen-Tannoudji} often discusses the fine structure of hydrogen by introducing perturbations such as the spin-orbit coupling ($\bm{S\cdot L}$),  the Darwin term ($\delta^3(\bm{r})$), and the relativistic correction ($\bm{p}^4$). These correspond to terms of order $1/M$ ($\psi^\dagger\bm{\sigma\cdot B}\psi$),  $1/M^2$  ($\psi^\dagger[\bm{\partial\cdot E}]\psi$), and  $1/M^3$ ($\psi^\dagger\bm{D}^4\psi$), in the NRQED Lagrangian,  respectively\footnote{The fine structure is calculated in quantum mechanics as a perturbation is $\alpha$, not in $1/M$, hence the mixture of terms of different orders in $1/M$.}. 

The original paper  \cite{Caswell:1985ui} has many citations. Since  this introduction is not meant as a comprehensive review, we will mention only a few relevant for our discussion. In a paper by Kinoshita and Nio \cite{Kinoshita:1995mt} the Feynman rules up to order $1/M^2$ (and some of the $1/M^3$ terms) were presented. The full $1/M^3$ terms in the NRQED Lagrangian were presented in a paper by Manohar \cite{Manohar:1997qy}. Also, the Wilson coefficients of the $1/M^3$ operators that contain one-photon coupling were calculated, for the case of a point particle,  to one-loop order \cite{Manohar:1997qy}. 

NRQED has been traditionally applied to point particles such as electrons and muons. Most of the applications of NRQED have focused on bound states of such particles. In the point particle case we know the full theory, QED, and the calculation of the Wilson coefficients is done in perturbation theory.  But NRQED can also be used to analyze the interaction of spin-half particle that is not elementary \cite{Hill:2011wy}.  The same NRQED Lagrangian is used, but the Wilson coefficients  are calculated non-perturbatively. For example, NRQED can provide a rigorous framework to analyze proton structure effects in systems like regular and muonic hydrogen \cite{Hill:2011wy}. In \cite{Hill:2012rh} the $1/M^4$ terms in the NRQED Lagrangian were presented and their Wilson coefficients were calculated. 

The purpose of this paper is to demonstrate how the NRQED Lagrangian is constructed. We will use general effective field theory arguments and will not assume that the spin-half field is elementary. The remainder of the paper is structured as follows. In section \ref{section2} we present the construction of the NRQED Lagrangian  up to and including order $1/M^2$. We explain how the Wilson coefficients are determined and their relation to the particle's charge, magnetic moment,  and  electric charge radius. We also comment on the relation between Heavy Quark Effective Theory (HQET) and NRQED. In section \ref{section3} we present a sample application, the calculation of the Thomson scattering differential cross section for photons off a spin-half particle. The result is familiar from classical electrodynamics and from the low energy limit of Compton scattering, but our derivation does not assume a spin-half point particle. We present our conclusions in section \ref{section4}.  

\section{Construction of the NRQED Lagrangian}\label{section2}
\subsection{General considerations} 
The NRQED Lagrangian describes the interaction of a spin-half non-relativistic field with an electromagnetic field. Since it is not manifestly Lorentz invariant\footnote{The fact that the entire Lagrangian is Lorentz invariant implies relations between the Wilson coefficients, see \cite{Manohar:1997qy, Heinonen:2012km,Hill:2012rh}.}, the Lagrangian can depend separately on the spatial $(A^i)$ and temporal $(A^0)$ components of the electromagnetic field. The Lagrangian is gauge and rotational invariant, Hermitian, and even under parity ($P$) and time reversal ($T$).  We discuss each property separately. 

\emph{Gauge invariance:} As a reminder, the relation between the electric $\bm E$  and magnetic $\bm B$ fields and the vector potential $\bm A$ and the scalar potential $A^0$ is $\bm{E}=-\partial \bm{A}/{\partial t}-\bm \nabla A^0$ and $\bm{B}=\bm{\nabla\times A}$. The electric and magnetic fields are invariant under the transformations $\bm A\to \bm A^\prime=\bm A-\bm\nabla \theta$, $A^0\to A^{\prime 0}=A^0+\partial \theta/\partial t$, where $\theta$ is an arbitrary  function of $\bm r $ and t. These are the time and space components of the Lorentz covariant gauge transformation $A^\mu\to A^{\prime \mu}=A^\mu+\partial^\mu \theta$. To enforce gauge invariance we define $D_t=\partial/\partial t+ieA^0$,  and $\bm D=\bm\nabla-ie\bm A$. These are the time and space components of the covariant derivative $D_\mu=\partial_\mu+ieA_\mu$. We assume\footnote{For a neutral spin-half particle see \cite{Hill:2012rh} } that the field $\psi$ has a charge $e$.  Under a gauge transformation $\psi\to e^{ie \theta(\bm r, t)}\psi $. As a result,  $D_t\psi\to e^{ie \theta(\bm r, t)}D_t\psi$ and  $\bm D\psi\to e^{ie \theta(\bm r, t)}\bm D\psi$, i.e. $D_t \psi$ and $\bm D\psi$ transform like $\psi$ under a gauge transformation. We will construct  the Lagrangian using these covariant derivatives. The physical meaning of the operators is often clearer if we express them in terms of the electric and magnetic fields. The electric and magnetic fields are related to the covariant derivatives by $E^j=\left(-i/e\right)[D_t,D^j]$ and $B^i=\epsilon^{ijk}\left(i/2e\right)[D^j,D^k]$. In constructing the Lagrangian we will use only the covariant derivatives, but we will often express various terms using the electric and magnetic fields.\\
\emph{Rotational  invariance:} Three-vector objects such as  $\bm D$, $\bm E$, $\bm B$, and $\bm \sigma$,  the Pauli matrices, are contracted with $\delta^{ij}$ and $\epsilon^{ijk}$ thus enforcing  rotational invariance.\\\emph{Hermiticity:} Hermiticity is enforced by using the Hermitian operators $iD_t$ and $i\bm D$ as building blocks and forming Hermitian combinations from products of these building blocks. \\
\emph{Parity:}  Since both $\bm \nabla$ and $\bm A$ are parity odd, $i\bm D$ is parity odd. Since both $\partial/\partial t$ and $A^0$  are parity even, $iD_t$ is parity even. The electric field $\bm E$ is parity odd while the magnetic field $\bm B$ is parity even.  Finally, $\bm\sigma$ is parity even. In the following we will use only parity even combinations of these objects. \\
\emph{Time reversal:}  Since both $i\bm\nabla$ and $\bm A$ are odd under time reversal, $i\bm D$ is odd under time reversal. Since both $i\partial/\partial t$ and $A^0$  are even under time reversal, $iD_t$ is  even under time reversal. The electric field $\bm E$ is even under time reversal,  while the magnetic field $\bm B$ is odd under time reversal.  Finally, $\bm\sigma$ is odd under time reversal. In the following we will use only time reversal even combinations of these objects.  \\

We can summarize the transformation properties above in the following table:
\begin{center}
\begin{tabular}{cccccc}
&$iD_t$&$i\bm D$&$\bm E$&$\bm B$&$\bm \sigma$\\
$P$&$+$&$-$&$-$&$+$&$+$\\
$T$&$+$&$-$&$+$&$-$&$-$
\end{tabular}
\end{center}
As a result, $P$ and $T$ invariance imply that we can only have even number of $i\bm D$'s, but any number of $iD_t$'s, when using the covariant derivatives as building blocks. We can use only one $\bm\sigma$, since $\sigma_i\sigma_j=\delta_{ij}+i\epsilon_{ijk}\sigma_k$.

In constructing the NRQED Lagrangian we assume that field redefinitions are made such that apart from the first term no factors of $iD_t$ are acting on the fermions. In other words,  $iD_t$ appears only in commutators. We will demonstrate this explicitly below. 

Ignoring four-fermion operators for now, the NRQED Lagrangian is of the form $${\cal L}=\sum_{n}\psi^\dagger \frac{O_n}{M^n}\psi,\quad n=0,1,2,\dots,$$
where $M$ is the mass of the fermion, e.g. electron, proton etc. Each $O_n$ is built out of covariant derivatives and their commutators, the electric and magnetic fields. The mass dimension of $\psi$ is $3/2$ and as a result the mass dimension of each $O_n$ is $n+1$. Note also that the mass dimension of $\bm\sigma$,  the covariant derivatives  $i\bm D$ and $iD_t$, and the electric and magnetic fields are zero, one, and two, respectively.   

\subsection{Construction of the leading power term}
The leading order term is of the form  $\psi^\dagger O_0\psi$. Since the dimension of $O_0$ is one, we can only construct it from one covariant derivative.  $i\bm D$ can only appear as $\bm{\sigma}\cdot \bm{iD}$ which is ruled out by parity, so we can only use $iD_t$. Thus we have a \emph{unique}\footnote{A ``mass" term of the form $M\psi^\dagger\psi$ can be eliminated using (\ref{Lagrangian0})  and the field redefinition $\psi\to e^{iMt}\psi$. The removal of the mass term is necessary to get a well defined power counting in 1/M.}  leading power term, $\psi^\dagger iD_t \psi$. We can normalize the coefficient of this term to be 1, so we have 
\begin{equation}\label{Lagrangian0}
{\cal L} = \psi^\dagger i D_t \psi +\cdots .
\end{equation}
\subsection{Construction of the $\bm{1/M}$ terms}
For the $1/M$ suppressed terms $O_1$ has dimension two. We can have two  $iD_t$'s, or two $i\bm{D}$'s. Terms that contain one $iD_t$ and one $i\bm{D}$ are ruled out by parity. 

Consider the first possibility, i.e. we have a term of the form $c\,\psi^\dagger \left(i D_t\right)^2 \psi/M$ in the Lagrangian, where $c$ is a real constant.  If we redefine $\psi$ via $\psi\to\psi -c\,iD_t\psi/2M$, the Lagrangian changes as
\begin{eqnarray*}
&& \psi^\dagger i D_t \psi +\dfrac{c}{M}\psi^\dagger \left(i D_t\right)^2 \psi\to \left(\psi -\dfrac{c}{2M}iD_t\psi\right)^\dagger \left(iD_t+\dfrac{c}{M}\left(i D_t\right)^2\right)\left(\psi -\dfrac{c}{2M}iD_t\psi\right)=\nonumber \\
 &=&\psi^\dagger i D_t \psi-\dfrac{c}{2M}\psi^\dagger(iD_t)^2\psi -\dfrac{c}{2M}\psi^\dagger(iD_t)^2\psi +\dfrac{c}{M}\psi^\dagger(iD_t)^2\psi +{\cal O}\left(\dfrac{1}{M^2}\right)=\nonumber \\
 &=& \psi^\dagger i D_t \psi +{\cal O}\left(\dfrac{1}{M^2}\right),
\end{eqnarray*}
eliminating the $1/M$ term \cite{Manohar:1997qy}. Notice that for any Hermitian operator $O$, $\psi^\dagger\left(iD_tO+OiD_t\right)\psi$ can be eliminated by an analogous field redefinition. 

Given two Hermitian operators $A$ and $B$ we can form two Hermitian combinations $AB+BA$ and $i(AB-BA)$. Similarly we can form two Hermitian operators $iD^iiD^j+iD^iiD^j$ and $i\sigma^k\left[iD^i,iD^j\right]$. We have to introduce $\sigma^k$ in the latter term to make it $T$-even. For the same reason we cannot multiply the former term by $\sigma^k$. The first combination has two indices and can only be contracted with $\delta^{ij}$, giving us the operator $\psi^\dagger{\bm D}^2\psi$.  The second has three indices and must be contracted with $\epsilon^{ijk}$, giving us the operator $i\psi^\dagger\epsilon^{ijk}\sigma^k\left[iD^i,iD^j\right]\psi$, which is proportional to $ \psi^\dagger\bm{\sigma\cdot B} \psi$. 

Allowing for arbitrary coefficients the NRQED Lagrangian up to order $1/M$  is 
\begin{equation}\label{Lagrangian1}
{\cal L} = \psi^\dagger\left( i D_t +c_2\dfrac{{\bm D}^2}{2M}+c_Fe\dfrac{\bm {\sigma\cdot B}}{2M}\right)\psi +\cdots .
\end{equation}
The subscript $F$ in $c_F$ stands for ``Fermi". The (hidden) Lorentz invariance of the Lagrangian implies that $c_2=1$ \cite{Manohar:1997qy, Heinonen:2012km,Hill:2012rh}.  To determine $c_F$ we need information concerning the one-photon interaction of the spin-half particle.  Such an interaction can be parametrized by the Dirac and Pauli form factors, $F_1$ and $F_2$, 
respectively,  defined by \cite{Foldy:1952,Salzman:1955zz}
$$
\langle \psi(p')|J_\mu^{\rm em}|\psi(p)\rangle=\bar u(p')
\left[\gamma_\mu F_1(q^2)+\frac{i\sigma_{\mu\nu}}{2M}F_2(q^2)q^\nu\right]u(p)\,,
$$
where $q^2=(p'-p)^2$. To calculate  the Wilson coefficients we follow the matching procedure of \cite{Manohar:1997qy}, but treating $F_1$ and $F_2$ as non-perturbative functions. We expand these functions in powers of $q^2/M^2$ and the spinors in  $\bm{p}/M$ and $\bm{p^\prime}/M$. Comparing the result to the NRQED calculation we find that for the leading term $F_1(0)$ gives the charge of $\psi$ in units of $e$. In particular, for our case we have $F_1(0)=1$. The $1/M$ terms give $c_2=1$, confirming the result above, and $c_F=F_1(0)+F_2(0)$ \cite{Manohar:1997qy, Hill:2012rh}. $F_2(0)$ is the anomalous magnetic moment of the particle. For a point particle, $F_2(0)=\alpha/2\pi+{\cal O}(\alpha^2)$.  The Wilson coefficients in (\ref{Lagrangian1}) therefore depend only on the charge and the anomalous magnetic moment of the spin-half particle.   

If we ignore the external electromagnetic field, the equations of motion imply that up to order $1/M$
$$
i\dfrac{\partial\psi}{\partial t}+\dfrac{\bm{\nabla}^2\psi}{2M}=0,
$$
which is of course Schr\"odinger equation. In other words, the NRQED Lagrangian is related to the Lagrangian formalism of non-relativistic quantum mechanics. 

\subsection{Construction of the $\bm{1/M^2}$ terms}
Construction of the ${1/M^2}$ Lagrangian is also fairly easy. Since we can only have an even number of  $i\bm{D}$'s we can have operators either with three $iD_t$'s or one $iD_t$ and two $i\bm{D}$'s. The first gives a unique operator $\psi^\dagger iD_t^3\psi$. This operator can be eliminated by a field redefinition similar to the elimination of $\psi^\dagger iD_t^2\psi$.  For the second option we can have six operators that are, suppressing the $\psi^\dagger$ ($\psi$) before (after) each term, $iD_tiD^iiD^j$,  $iD_tiD^jiD^i$, $iD^iiD^jiD_t$, $iD^jiD^iiD_t$, $iD^iiD_tiD^j$, $iD^jiD_tiD^i$. 

From the first four we can form  Hermitian combinations by taking the sum and $i\sigma^k$ times the difference of each Hermitian conjugate pair.  The former must be contracted with $\delta^{ij}$ and the latter with $\epsilon^{ijk}$. We get two Hermitian combinations,
\begin{eqnarray*}
&&\delta^{ij}\left(iD_tiD^iiD^j+iD^jiD^iiD_t\right)=-\left(iD_t{\bm D}^2+{\bm D}^2iD_t\right), \\
&&\epsilon^{ijk}i\sigma^k\left(iD_tiD^iiD^j-iD^jiD^iiD_t\right)=\epsilon^{ijk}i\sigma^k\left(iD_tiD^iiD^j+iD^iiD^jiD_t\right)=\\&&=\dfrac12\epsilon^{ijk}i\sigma^k\left(iD_t\left[iD^i,iD^j\right]+\left[iD^i,iD^j\right]iD_t\right)=-\left(iD_te\bm{\sigma\cdot B}+e\bm{\sigma\cdot B}iD_t\right).
\end{eqnarray*}
Both terms are of the form $\left(iD_tO+OiD_t\right)$ and can be eliminated using a field redefinition.  

We are left with the operators $iD^iiD_tiD^j$ and $iD^jiD_tiD^i$. By commuting $iD_t$ to the left or to the right we can write, for example, $iD^iiD_tiD^j$ as $[iD^i,iD_t]iD^j+iD_tiD^iiD^j$ or $iD^i[iD_t,iD^j]+iD^iiD^jiD_t$. We have already analyzed terms in which $iD_t$ is on the leftmost or rightmost position.  The commutator is proportional to the electric field $\bm E$, so we can consider operators constructed out of $iD^i$ and $E^j$ instead of $iD^iiD_tiD^j$ and  $iD^jiD_tiD^i$.

From the operators $iD^i$ and $E^j$ we can form two Hermitian combinations: $\sigma^k\left(iD^iE^j+E^jiD^i\right)$ and $i[iD^i,E^j]$. Since $iD^iE^j$ and $E^jiD^i$ are $T$-odd, the $\sigma^k$ must appear with symmetric combination.  Contacting $\sigma^k\left(iD^iE^j+E^jiD^i\right)$  with $\epsilon^{ijk}$ and $i[iD^i,E^j]$ with $\delta^{ij}$ we find two operators, 
\begin{eqnarray*}
&&\epsilon^{ijk}\sigma^k\left(iD^iE^j+E^jiD^i\right)=\epsilon^{ijk}\sigma^k\left(iD^iE^j-E^iiD^j\right)=i\bm{\sigma}\cdot\left(\bm{D\times E}-\bm{E\times D}\right),  \\
&&\delta^{ij}i[iD^i,E^j]=-\bm{\partial\cdot E}.
\end{eqnarray*}
Allowing for arbitrary coefficients the NRQED Lagrangian up to order $1/M^2$  is 
\begin{equation}\label{Lagrangian2}
{\cal L} = \psi^\dagger\left\{ i D_t +c_2\dfrac{{\bm D}^2}{2M}+c_Fe\dfrac{\bm {\sigma\cdot B}}{2M}+c_De\dfrac{[\bm{\partial\cdot E}]}{8M^2}+ic_Se\dfrac{\bm{\sigma}\cdot\left(\bm{D\times E}-\bm{E\times D}\right)}{8M^2}\right\}\psi +\cdots .
\end{equation}
The notation $[\bm{\partial\cdot E}]$ denotes that the derivative is acting only on $\bm E$ and not on $\psi$. The subscript $D$ in $c_D$ stands for ``Darwin" \cite{Darwin} and the subscript $S$ in $c_S$ stands for ``Seagull" \cite{Caswell:1985ui,Kinoshita:1995mt}. For a classical electric field the Darwin term gives rise to the $\delta(\bm{r}^3)$ perturbation for the hydrogen spectrum familiar from non-relativistic quantum mechanics.  Similar to $c_2$ above, $c_S$ is determined from the (hidden) Lorentz invariance of the full Lagrangian, $c_S=2c_F-F_1(0)$ \cite{Manohar:1997qy, Heinonen:2012km,Hill:2012rh}. $c_D$ is determined in a similar way to $c_F$, i.e. via matching. Performing the matching at order $1/M^2$ gives $c_S=2c_F-F_1(0)$, confirming the result above and $c_D=F_1(0)+2F_2(0)+8M^2F_1'(0)$, where $F_1^\prime=dF_1(q^2)/dq^2$. $F_1^\prime(0)$ is related to the definition of the electric charge radius of the spin-half particle.

At this order the operators appearing in the Lagrangian are of mass dimension six. As a result we can write operators that couple four spin-half fields. There are two possible operators, a spin-dependent and a spin-independent operator, 
 
\begin{equation}\label{Lagrangian4f}
{\cal L}_{\psi\chi}=\dfrac{d_1}{M^2}\psi^\dagger\sigma^i\psi\chi^\dagger\sigma^i\chi+\dfrac{d_2}{M^2}\psi^\dagger\psi\chi^\dagger\chi+\cdots .
\end{equation}
Here $\chi$ is another NRQED field which \emph{can} be different\footnote{Other structures, e.g. $\psi^\dagger\chi\chi^\dagger\psi$, can be related to (\ref{Lagrangian4f}) via Fierz-type identities, see e.g. \cite{Nishi:2004st}.} from $\psi$. We follow the notation of \cite{Hill:2012rh}. The notation for the coefficients follows that of  \cite{Kinoshita:1995mt}. In application of NRQED $\chi$ is often an electron or a muon. To determine the coefficients $d_1$ and $d_2$ one has to do a matching calculation that is more involved than that of $c_i$, see \cite{Hill:2011wy}.

\subsection{Order $\bm{1/M^3}$ and beyond}

The formalism presented above can be extended to order $1/M^3$ \cite{Manohar:1997qy} and $1/M^4$ \cite{Hill:2012rh}. The number of operators starts to proliferate as we go to higher orders in $1/M$. Thus we find  for operators of the form $\psi^\dagger\dots \psi$,  seven terms at order $1/M^3$ \cite{Manohar:1997qy}, and twelve terms at order $1/M^4$ \cite{Hill:2012rh}. In constructing these terms it is more convenient to use both covariant derivatives and the electric and magnetic fields as building blocks.  Also, at higher orders one needs to apply new constraints such as the homogenous Maxwell equations  $[\bm{\partial\cdot B}]=0$, and $[\partial _t\bm{B}+\bm{\partial\times E}]=0$, see \cite{Hill:2012rh} for details.  At order $1/M^4$ there are new contributions to ${\cal L}_{\psi\chi}$, and new pure photonic operators that can be identified with the Euler-Heisenberg Lagrangian \cite{Hill:2012rh}. At order $1/M^3$ and higher one first encounters operators that only couple to two photons or more. To determine the values of their Wilson coefficients requires the knowledge of the low energy limit of Compton scattering, with real or virtual photons  \cite{Hill:2012rh}.  

\subsection{The relation between HQET and NRQED}
As discussed in detail in \cite{Manohar:1997qy}, there is a close connection between Heavy Quark Effective Theory (HQET) and Non-Relativistic QCD (NRQCD) Lagrangians. Replacing the gauge group by $U(1)_{\scriptsize\mbox{EM}}$ give a similar relation to NRQED. We can thus take the QED Lagrangian and follow a well-known procedure, see e.g. \cite{Manohar:2000dt}, to obtain  
$$
{\cal L}_v=\bar\psi_v\left(iv\cdot D+i\Dslash_\perp\dfrac1{2M+iv\cdot D}i\Dslash_\perp\right)\psi_v,
$$
where $\psi_v(x)=e^{iMv\cdot x}\left(1+\vslash\right)\psi/2$ and $D_\perp^\mu=D^\mu-v^\mu v\cdot D$. If we take $v=(1,0,0,0)$ we have $iv\cdot D=iD_t$ and $D_\perp^\mu\to D^i$.  Expanding up to order $1/M^2$ we obtain
$$
{\cal L}_v=\psi_v^\dagger\left(iD_t+\dfrac1{2M}\gamma^i\gamma^jiD^iiD^j-\dfrac1{4M^2}\gamma^i\gamma^jiD^iiD_tiD^j+\cdots \right)\psi_v.
$$
 We recognize the same operators we have analyzed above. The anti commutator of  $\gamma^i$$\gamma^j$ is proportional to $\delta^{ij}$, while $\left(1+\vslash\right)[\gamma^i,\gamma^j]\left(1+\vslash\right)$ is proportional to $\epsilon^{ijk}\sigma^k$. With the help of a field redefinition we can bring it to the form of (\ref{Lagrangian2}) with the result $c_2=c_F=c_D=c_S=1$. These are the tree level values, while the one-loop expressions can be found in \cite{Kinoshita:1995mt, Manohar:1997qy}.  
 
 This gives a direct relation between the QED point-particle, HQET, and NRQED Lagrangians. Notice though that it is less general than the procedure described above that applies also to a composite spin-half particle.  In fact, already at order $1/M^3$ the expansion of the point particle Lagrangian will not generate all the possible operators at this order\footnote{In other words, some of the $1/M^3$ operators have zero Wilson coefficients at tree level.} \cite{Manohar:1997qy}.  Although the Lagrangians look the same, the power counting for each theory is different. In particular the fermion propagators for HQET and NRQED are different, see \cite{Manohar:1997qy} for details.

\section{Sample application: Thomson scattering for a 
possibly composite
spin-half particle}\label{section3}  
As a sample application of NRQED, we calculate the Thomson scattering cross section of a photon off a spin-half particle. The particle can be elementary or composite. The answer will depend only on the mass and  total charge of the particle.   We are therefore calculating the low photon energy  limit of the cross section for $\psi(p)+\gamma(q)\to\psi(p^\prime)+\gamma(q^\prime)$ at the rest frame of $\psi$ at the lowest order in $1/M$. At this order we have $p=p^\prime=(M,\bm{0}), q^0=q^{\prime0}$, and $|\bm{q}|=|\bm{q^\prime}|$.

The lowest term in the Lagrangian (\ref{Lagrangian0}) couples fermions and time-like photons. As a result, it does not contribute to real photon scattering.  At order $1/M$ we have two operators, see (\ref{Lagrangian1}), but only one of them couples to two real photons at order  $1/M$.  We have 
$$
{\cal L}\ni\psi^\dagger\dfrac{{\bm D}^2}{2M}\psi=\dfrac1{2M}\psi^\dagger\left(\dfrac{\partial}{\partial x^i}-ieA_i\right)^2\psi\Rightarrow -\dfrac{e^2}{2M}\psi^\dagger \bm{A\cdot A}\psi \Rightarrow -\dfrac{e^2}{2M}\bar \psi\gamma^0 \delta^{ij}A^iA^j\psi +\cdots .
$$
To simplify the presentation, we have reverted to the usual Dirac fields in the last step which is justified at this order.  The resulting Feynman rule is $-2i\dfrac{e^2}{2M}\gamma^0\delta^{ij}$. The amplitude is 
$$
{\cal M}=-\dfrac{e^2}{M}\bm{\epsilon(q)\cdot \epsilon^*(q^\prime)}\, \bar u(p)\gamma^0u(p),
$$
and the spin averaged amplitude is 
$$
|\overline{{\cal M}}|^2=\dfrac14\dfrac{e^4}{M^2}
\sum_{\scriptsize\mbox{pol.}}|\bm{\epsilon(q)\cdot \epsilon^*(q^\prime)}|^2
\mbox{Tr}\left[\left(\pslash+M\right)\gamma^0\left(\pslash+M\right)\gamma^0\right]=2e^4\sum_{\scriptsize\mbox{pol.}}|\bm{\epsilon(q)\cdot \epsilon^*(q^\prime)}|^2,
$$
where we have used $\pslash=M\gamma^0$. For a real photon we have $\sum_{\scriptsize\mbox{pol.}}\epsilon_i(\bm q)\epsilon^*_j(\bm q)=\delta_{ij}-\hat{q}_i\hat{q}_j$, where $\bm{\hat q}=\bm{ q}/|\bm{ q}|$ \cite{Landau}. We have 
$$
|\overline{{\cal M}}|^2=2e^4\left(\delta_{ij}-\hat{q}_i\hat{q}_j\right)\left(\delta_{ij}-\hat{q}_i^{\,\prime}\hat{q}_j^{\,\prime}\right)=2e^4\left[1+\left(\bm{\hat{q}\cdot\hat{q}^{\,\prime}}\right)^2\right]=2e^4\left(1+\cos^2\theta\right),
$$
where $\theta$ is the angle between the incoming ($\bm q$) and outgoing ($\bm q^\prime$)  photon momenta, i.e. the   scattering angle. 
In the large $M$ limit the spin averaged cross section is 
$$
\dfrac{d\sigma}{d\Omega}=\left(\dfrac1{8\pi M}\right)^2|\overline{{\cal M}}|^2=\dfrac{\alpha^2}{2M^2}\left(1+\cos^2\theta\right).
$$
The total cross section is 
$$\sigma=\dfrac{8\pi}{3}\left(\dfrac{\alpha}{M}\right)^2,
$$
which is of course the Thomson cross section \cite{Thomson}. This result applies for the low energy limit of the scattering of a photon for any spin-half fermion with charge $\pm e$, elementary or composite.

We can easily calculate the higher order corrections in $1/M$ to the scattering by using NRQED Feynman rules. In particular $1/M^2$ terms in the amplitude $\cal M$ arise from  tree level diagrams that involve two insertions of (\ref{Lagrangian1}) and one insertion of (\ref{Lagrangian2}). These results reproduce the well known low-energy theorems of \cite{Low:1954kd} and \cite{GellMann:1954kc}, but the derivation is much simpler. For example, \cite{GellMann:1954kc} carefully expands Green's functions to obtain the amplitude. Using NRQED we only need to calculate a few tree level Feynman diagrams. The final answer depends on the Wilson coefficients\footnote{$c_D$ does not contribute at this order since the operator does not couple to real photons.} $c_F$ and $c_S$ which depend in turn on $F_1(0)$ and $F_2(0)$, i.e. the charge and magnetic moment of the spin-half particle. In fact, it is easy to reach this conclusion without even doing any calculation, illustrating again the power of NRQED.   

\section{Conclusions}\label{section4}

We have presented a pedagogical introduction to Non Relativistic QED (NRQED), the effective field theory that describes the interaction of the electromagnetic fields with a non-relativistic possibly composite spin-half particle. NRQED itself was first formulated as an effective field theory 30 years ago, but its roots can be traced to the early days of quantum mechanics. Traditionally NRQED has been applied to bound QED problems, but it can be applied to problems that involve composite particles such as the proton and the neutron. Only fairly recently such applications of NRQED were considered in the literature.   

We have explicitly described how one derives the leading order term, the $1/M$ terms, and the $1/M^2$ terms in the NRQED Lagrangian. The $1/M^3$ \cite{Manohar:1997qy} and $1/M^4$ \cite{Hill:2012rh} terms in the NRQED Lagrangian are also known, and can be derived in a similar way.  We have also commented on the relation between the NRQED and the HQET Lagrangians.

As a sample application, we have presented the derivation of the well-known Thomson cross section. The result applies for the low energy scattering of a real photon on \emph{any} charged spin-half particle, composite or elementary. It is easy to derive $1/M$ power corrections to the cross section by calculating higher power NRQED tree level diagrams. 

Throughout this paper we have considered the parity and time-reversal even interaction of a  spin-half particle coupled to an abelian gauge field. Similar effective field theories can be constructed for other particle spins, for non-abelian gauge symmetry (NRQCD), and for $P$ and $T$ violating interactions.  

\vskip 0.2in
\noindent
{\bf Acknowledgements}
\vskip 0.1in
\noindent
A large potion of the material presented here was a direct result of my collaboration with Richard J. Hill, Gabriel Lee, and Mikhail P. Solon. I thank them for it and for their comments on the manuscript. I thank Alexey A. Petrov for encouraging me to write up this introduction to NRQED and for his comments on the manuscript.  I  also thank Christopher Brust for a useful discussion. I thank  Perimeter Institute for Theoretical Physics for its hospitality and support during the completion of this work. This work is supported by the  DOE grant DE-FG02-13ER41997 and the NIST Precision Measurement Grants Program.


\begin{thebibliography}{99}

 \bibitem{Caswell:1985ui}
  W.~E.~Caswell and G.~P.~Lepage,
  %``Effective Lagrangians For Bound State Problems In QED, QCD, And Other Field
  %Theories,''
  Phys.\ Lett.\  B {\bf 167}, 437 (1986).
  %%CITATION = PHLTA,B167,437;%%

\bibitem{Cohen-Tannoudji} 
Claude Cohen-Tannoudji, Bernard Diu, and Frank Lalo', ``Quantum Mechanics" Vol. II, Wiley-Interscience (1977)

\bibitem{Kinoshita:1995mt}
  T.~Kinoshita and M.~Nio,
  %``Radiative Corrections to the Muonium Hyperfine Structure. I. The $\alpha~2
  %(Z\alpha)$ Correction,''
  Phys.\ Rev.\  D {\bf 53}, 4909 (1996).
  [arXiv:hep-ph/9512327].
  %%CITATION = PHRVA,D53,4909;%% 
  
\bibitem{Manohar:1997qy} 
  A.~V.~Manohar,
  %``The HQET / NRQCD Lagrangian to order alpha / m-3,''
  Phys.\ Rev.\ D {\bf 56}, 230 (1997)
  [hep-ph/9701294].
  
\bibitem{Hill:2011wy} 
  R.~J.~Hill and G.~Paz,
  %``Model independent analysis of proton structure for hydrogenic bound states,''
  Phys.\ Rev.\ Lett.\  {\bf 107}, 160402 (2011)
  [arXiv:1103.4617 [hep-ph]].
  %%CITATION = ARXIV:1103.4617;%%
  
 \bibitem{Hill:2012rh} 
  R.~J.~Hill, G.~Lee, G.~Paz and M.~P.~Solon,
  %``NRQED Lagrangian at order $1/M^4$,''
  Phys.\ Rev.\ D {\bf 87}, no. 5, 053017 (2013)
  [arXiv:1212.4508 [hep-ph]].


\bibitem{Heinonen:2012km} 
  J.~Heinonen, R.~J.~Hill and M.~P.~Solon,
  %``Lorentz invariance in heavy particle effective theories,''
  Phys.\ Rev.\ D {\bf 86}, 094020 (2012)
  [arXiv:1208.0601 [hep-ph]]
  
\bibitem{Foldy:1952}
  L.~L.~Foldy,
  %``The Electromagnetic Properties  of Dirac Particles,''
  Phys.\ Rev.\  {\bf 87}, 688 (1952).
  %%CITATION = PHRVA,87,688;%%

\bibitem{Salzman:1955zz}
  G.~Salzman,
  %``Neutron-Electron Interaction,''
  Phys.\ Rev.\  {\bf 99}, 973 (1955).
  %%CITATION = PHRVA,99,973;%% 
  
\bibitem{Darwin} 
   C.~G.~Darwin,
  %``The Wave Equations  of the Electron,''
  Proc.\ Roy.\ Soc.\ Lond.\ A {\bf 118}, 654 (1928).
  
 \bibitem{Nishi:2004st} 
 C.~C.~Nishi,
  %``Simple derivation of general Fierz-like identities,''
  Am.\ J.\ Phys.\  {\bf 73}, 1160 (2005)
  [hep-ph/0412245]. 
  
  
\bibitem{Manohar:2000dt} 
  A.~V.~Manohar and M.~B.~Wise,
  ``Heavy quark physics,''
  Camb.\ Monogr.\ Part.\ Phys.\ Nucl.\ Phys.\ Cosmol.\  {\bf 10}, 1 (2000).
  %%CITATION = CMPCE,10,1;%%  
  
   
  
\bibitem{Landau}
 L.D. Landau and E.M. Lifshitz, 
`` The Classical Theory of Fields," (4th ed.) 
 Butterworth-Heinemann (1975).
  
\bibitem{Thomson}
 J.J. Thomson, 
 ``Conduction of electricity through gases," (2nd ed.)
 p. 325,  
Cambridge University Press  (1906).  
  
  
  
\bibitem{Low:1954kd}
  F.~E.~Low,
  %``Scattering of light of very low frequency by systems of spin 1/2,''
  Phys.\ Rev.\  {\bf 96}, 1428-1432 (1954).
  
 \bibitem{GellMann:1954kc}
  M.~Gell-Mann, M.~L.~Goldberger,
  %``Scattering of low-energy photons by particles of spin 1/2,''
  Phys.\ Rev.\  {\bf 96}, 1433-1438 (1954). 
  
  
  

\end{thebibliography}
\end{document}